\documentclass[10pt,leqno]{amsart}
\usepackage{graphicx}
\usepackage{indentfirst,csquotes}
\usepackage{amssymb,amsthm,amsmath}
\usepackage{xcolor,paralist,hyperref,fancyhdr,etoolbox}
\usepackage{multirow,booktabs}

\hypersetup{ colorlinks=true, linkcolor=black, filecolor=black, urlcolor=black }

\begin{document}

\title[BeeVe: Unsupervised Acoustic State Discovery in Honey Bee Buzzing]
{BeeVe: Unsupervised Discovery of Non-Semantic Acoustic States, Towards a Non-Invasive Assessment of Honey Bee Colony Health}

\author[H. Hammami]{Hamze Hammami}
\address{School of Engineering and Physical Sciences, Heriot-Watt University Dubai}
\email{hh2095@hw.ac.uk}

\author[N. Abdulaziz]{Nidhal Abdulaziz}
\address{School of Engineering and Physical Sciences, Heriot-Watt University Dubai}
\email{Nidhal.Abdulaziz@hw.ac.uk}

\date{\today}

\maketitle
\let\thefootnote\relax
\footnotetext{MSC2020: Primary 68T07 (Machine learning), Secondary 92B05 (General biology).}

\begin{abstract}
Discovering structure in biological signals without supervision is a fundamental problem in computational intelligence, yet existing bioacoustic methods assume vocal production models or predefined semantic units, leaving non-vocal species poorly served. Honey bees are a compelling instance of this gap: their collective buzzing arises from mechanical muscle vibrations rather than any communicative vocal apparatus, and while evidence suggests these vibrations reflect colony physiological state, no existing vocal framework applies. This work introduces BeeVe, an unsupervised framework for acoustic state discovery in collective honey bee buzzing. BeeVe uses the self-supervised Patchout Spectrogram Transformer (PaSST) as a frozen feature extractor to produce general acoustic embeddings, then trains a Vector-Quantized Variational Autoencoder (VQ-VAE) entirely without labels on those embeddings, learning a finite discrete codebook of acoustic tokens 
directly from unlabelled hive audio. All learning applied to bee audio is unsupervised: no labels, pretext tasks, or contrastive objectives are used at any stage. Each token represents a recurring acoustic pattern, and the full codebook forms a reusable vocabulary of colony-level acoustic states discovered entirely without annotation. Post-hoc evaluation against known queen status reveals that the learned tokens separate queenright and queenless conditions with Jensen-Shannon Divergence values between 0.609 and 0.688, and that the queenless condition further decomposes into three internally coherent sub-states stable across experiments with different codebook sizes and random seeds. Token transition analysis further confirms non-random sequential structure in the learned vocabulary ($p \ll 0.001$), consistent across all three experiments. Generalisation to unseen recordings preserves both token overlap (Jaccard = 0.947) and global manifold topology. These results demonstrate that unsupervised discrete codebook learning can recover repeatable acoustic structure from a non-vocal biological signal without annotation, opening a path toward non-invasive acoustic hive health monitoring.
\end{abstract}

\bigskip
\noindent\textbf{Keywords:} bioacoustics, PaSST, representation learning, state discovery, unsupervised learning, VQ-VAE.

$\,$

\section{Introduction}

Honey bees are essential pollinators whose global decline
poses serious risks to ecosystems dependent on their pollination. While bee
communication is primarily understood through pheromonal signals and the
waggle dance \cite{michelsen1986sound}, evidence suggests that collective
buzzing reflects deeper physiological and behavioural states of the colony.
Acoustic correlates have been identified for queen presence and loss
\cite{kirchner1993acoustical, kanelis2023decoding}, swarming preparation
\cite{ferrari2008monitoring}, and modulatory vibrational signals such as the
dorso-ventral abdominal vibration \cite{ramsey2018extensive}, indicating that
the acoustic output of a hive is not random noise but structured emergent
behaviour linked to internal colony state. This makes acoustic monitoring a
promising non-invasive alternative to physical hive inspection, with early
detection of conditions such as queen loss or swarming offering practical
benefits for colony health management \cite{abdollahi2022automated}.

Building on the acoustic monitoring direction, Hammami and Abdulaziz \cite{hammami2024beebetter} proposed a supervised classifier for queen
status detection from hive audio, demonstrating that acoustic signals are
sufficiently structured for machine learning-based state detection. The
present work extends this by removing the supervision requirement entirely,
asking whether that structure can be discovered without any labels.

Early attempts to record animal sounds date to the late 19th century \cite{erbe2022exploring}.
A significant landmark demonstration specific to bees came from Karl von Frisch
\cite{frisch1993dance}, whose decoding of the waggle dance showed that bee
behaviour encodes structured spatial information, earning the Nobel Prize in
1973 and establishing that signals produced by bees carry meaning beyond
simple reflex.

Modern non-human communication research has advanced significantly through
machine learning and representation learning, with large-scale models
increasingly explored as tools for analysing animal acoustic systems.
These advances have enabled self-supervised approaches to pattern discovery in animal vocalisations, learning meaningful representations directly from unlabelled audio without manual annotation, particularly valuable where the
underlying structure and semantics remain unknown.

NatureLM-Audio \cite{robinson2024naturelm} presents an audio-language foundation model that learns general acoustic features from unlabelled animal sound data through self-supervised representation learning. This addresses a key limitation of earlier studies, namely the reliance on narrowly curated datasets and task-dependent feature engineering, and produces representations transferable across species and tasks.

Building on this idea, AVES \cite{hagiwara2023aves} introduced a
self-supervised transformer encoder for animal vocalizations, demonstrating
that meaningful acoustic representations can be learned without manual
annotation and transferred effectively across species and tasks, showing that
latent representations learned from animal sounds exhibit structure that is
useful for downstream tasks. A direct investigation of this transferability
was provided by \cite{sarkar2025comparing}, who systematically compared SSL
models pre-trained on animal vocalizations, specifically AVES, against models
pre-trained on human speech, across three bioacoustic datasets. Their results
showed that pre-training on bioacoustic data yields marginal improvements over
speech-pretrained models in most settings, and that further fine-tuning on
automatic speech recognition tasks produces inconsistent gains. A significant
finding suggests that the general-purpose representations capture structure
sufficiently transferable to non-human vocalizations.

The WhaleLM framework \cite{sharma2024whalelm} applies neural sequence models
to sperm whale vocalizations and demonstrates that whale codas exhibit
long-range dependencies and structure. The study shows that these vocal
sequences encode information about both current and future behavior, providing
evidence that animal communication systems may contain higher-order
structure. Importantly, this work operates on coda unit types already identified 
and segmented by marine biologists, meaning the discrete units are 
externally defined rather than learned from raw audio. Complementing this
line of work, Paradise et al.\ introduced WhAM \cite{paradise2025wham}, which
models sperm whale vocalizations using translation neural architectures. Their
approach frames animal communication as a representation learning problem,
where latent embeddings capture communicative structure without explicit
human-defined labels.

A particularly relevant line of work concerns the discretization of animal
vocalizations into sequences of learned acoustic tokens, a concept very much
aligned with this study. \cite{sarkar2025towards} investigated if discrete
token sequences through vector quantization of SSL embeddings can capture and
leverage the temporal structure of animal calls, applying this framework to
four datasets covering marmosets and dogs. Their distance analysis
demonstrated that vector-quantized token sequences generated from HuBERT
embeddings exhibit meaningful separability by call-type and caller identity,
and that a k-nearest neighbour classifier operating on Levenshtein distances
between token sequences achieves reasonable classification performance. This
represents the first application of discrete audio tokenization to
bioacoustics, establishing that learned codebooks can capture relevant
variation in animal vocalizations, a finding directly motivating the approach
taken in the present work.

Further supporting the view that non-human communication can be studied using modern ML algorithms, a very recent study by James et al.\
\cite{james2026zebra} extended a discrete acoustic tokenization to a real-time interactive setting, developing an audio-LLM that engages in naturalistic vocal exchanges with live zebra finches. Their system represents individual calls as discrete tokens learned through vector quantization, demonstrating that compact learned codebooks can capture meaningful acoustic variation in animal vocalizations and enable downstream sequence modeling.

Carvalho~Jr.\ et al.\ \cite{carvalho2025unsupervised} apply a convolutional autoencoder with HDBSCAN clustering to bee audio for unsupervised queenless 
detection, demonstrating that unsupervised methods can match or exceed supervised baselines on this task. However, the approach targets binary 
classification and produces no reusable discrete vocabulary of acoustic states.

These advances share a common principle, structure in animal acoustics 
can be recovered from data without a predefined semantic assumptions. 
However, all prior work targets \textit{vocal} species, those producing 
sound through a dedicated phonatory apparatus \cite{fitch2000evolution}, 
 and mostly relies on predefined discrete units, as in WhaleLM and 
WhAM. Honey bee buzzing is \textit{non-vocal}, arising from muscle vibrations reflecting the collective  state rather than communicative intent. Table~\ref{tab:approach_comparison} positions key 
works across signal type, supervision, and discretization.

\begin{table}[!t]
\renewcommand{\arraystretch}{1.3}
\caption{Comparison of Bioacoustic Approaches}
\label{tab:approach_comparison}
\centering
\small
\begin{tabular}{lccc}
\toprule
\textbf{Method} & \textbf{Signal} & \textbf{Supervision} & \textbf{Learned Codebook} \\
\midrule
NatureLM \cite{robinson2024naturelm}                 & Vocal     & Self-sup.  & --         \\
AVES \cite{hagiwara2023aves}                          & Vocal     & Self-sup.  & --         \\
WhaleLM \cite{sharma2024whalelm}                      & Vocal     & Self-sup.  & --         \\
WhAM \cite{paradise2025wham}                          & Vocal     & Self-sup.  & --         \\
Sarkar \& Doss \cite{sarkar2025towards}               & Vocal     & Self-sup.  & \checkmark \\
James et al.\ \cite{james2026zebra}                   & Vocal     & Unsup.     & \checkmark \\
Abdollahi et al.\ \cite{abdollahi2022automated}       & Non-vocal & Supervised & --         \\
Kanelis et al.\ \cite{kanelis2023decoding}            & Non-vocal & Supervised & --         \\
Carvalho Jr.\ et al.\ \cite{carvalho2025unsupervised} & Non-vocal & Unsup.     & --         \\
\textbf{BeeVe }                                 & \textbf{Non-vocal} & \textbf{Unsup.} & \checkmark \\
\bottomrule
\end{tabular}
\end{table}

BeeVe frames the signal as non-vocal and applies unsupervised discrete codebook learning directly to bee audio, where PaSST serves as a frozen feature extractor pretrained on AudioSet \cite{koutini2021efficient} and queen status is used only for post-hoc evaluation.

Accordingly, this work does not model honey-bee buzzing as a language, nor does it aim to decode explicit communicative messages. Instead, a state-discovery perspective is adopted where semantic units, communicative
intent, and generative structure are not assumed. Representations emerge through unsupervised learning, and the resulting structure is evaluated against known states and examined for finer sub-state organisation. The central question is whether non-random, compositional structure is inherent 
in collective honey-bee buzzing as an emergent property of colony behaviour. 
This work frames that question as a computational intelligence problem, applying unsupervised representation learning and discrete codebook learning to recover structure from a biological signal without annotation.

The main contributions of this work are:
\begin{itemize}
    \item Proposes a framework for unsupervised discrete acoustic pattern discovery applied to a non-vocal species, making no semantic assumptions about the signal.
    \item Experimental demonstration that VQ-VAE tokens separate queenright and queenless state conditions (JSD 0.609-0.688) without any label supervision.
    \item Characterisation of three internally stable queenless sub-states consistent across different codebook sizes and random seeds, suggesting the possibility of distinct behavioural modes within the queenless condition.
    \item Statistical evidence of non-random temporal structure in the learned token sequences. 
\end{itemize}

\section{Methodology}

\begin{figure}[!htbp]
    \centering
    \includegraphics[width=\columnwidth]{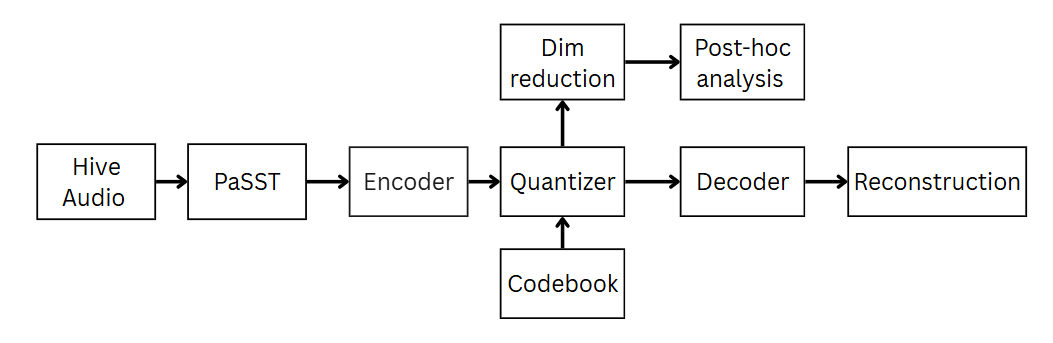}
    \caption{Method summary.}
    \label{fig:method_summary}
\end{figure}

The block diagram in Figure~\ref{fig:method_summary} summarises the general methodology.

\subsection{UrBAN Dataset}

Acoustic data equivalent to approximately five hours of hive recordings was
sampled from the UrBAN dataset \cite{abdollahi2025urban}. The UrBAN dataset
contains well over 1000 hours of honey-bee hive audio collected under
real-world conditions. For the purposes of a controlled and interpretable
experiment, the amount of data used is limited while preserving data diversity
through annotations. These annotations are used only to ensure that recordings
from multiple conditions are represented in the experiment and are not used as
labels for training.

\subsection{Feature Extraction and Model Architecture}

Audio data is transformed into acoustic representations using the Patchout Spectrogram Transformer (PaSST), audio is loaded at 22{,}050\,Hz before feature extraction. The model variant \texttt{passt\_s\_swa\_p16\_128\_ap476} is used, loaded from the \textit{hear21passt} library \cite{koutini2021efficient} with pretrained weights and no fine-tuning. Timestamp embeddings are extracted at approximately 23\,ms intervals (hop length 512 at 22{,}050\,Hz), yielding one 1295-dimensional feature vector per frame.

\begin{figure}[!htbp]
    \centering
    \includegraphics[width=0.7\columnwidth]{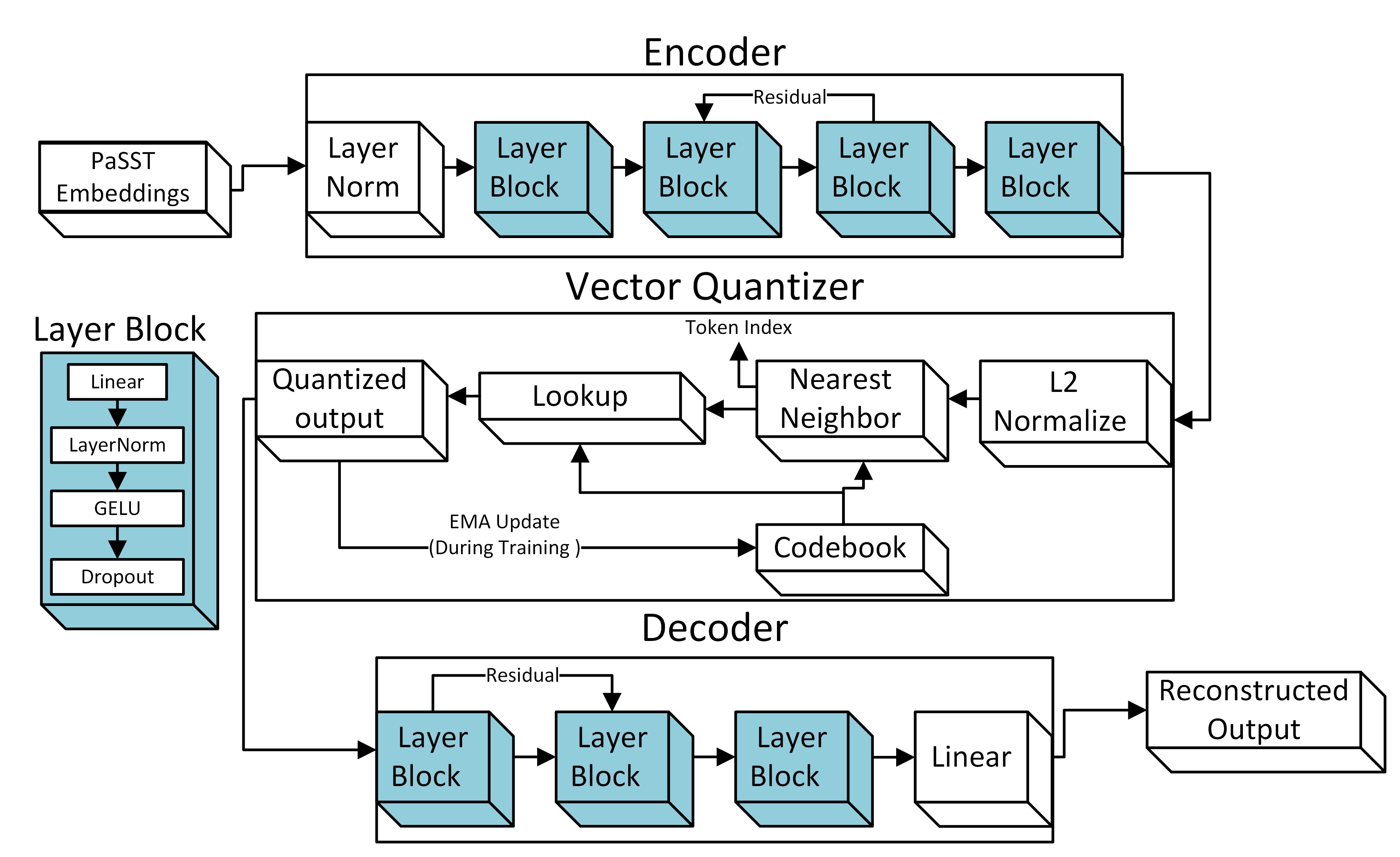}
    \caption{VQ-VAE architecture.}
    \label{fig:vqarch}
\end{figure}

The feature vector embeddings are passed as the input to a Vector-Quantized Variational Autoencoder (VQ-VAE) the exact architecture can be seen in Figure~\ref{fig:vqarch}, consisting of three components:

\begin{itemize}
    \item \textbf{Encoder:} Compresses PaSST embeddings into
    a 128-dimensional continuous latent representation $\mathbf{z}_e \in\mathbb{R}^{128}$ through five fully connected blocks
    ($1295 \rightarrow 1024 \rightarrow 512 \rightarrow 512 \rightarrow 128$)with LayerNorm, GELU activation, dropout, and a residual connection at the $512 \rightarrow 512$ stage.
    \item \textbf{Vector Quantizer:} Maps the latent representation $\mathbf{z}_e$ to the nearest entry in a learned codebook $\mathcal{C} = \{\mathbf{e}_1, \mathbf{e}_2, \ldots, \mathbf{e}_K\}$, where $K \in \{32, 64\}$ depending on the experiment, $K=64$ is the primary baseline at this data scale (five hours), $K=32$ is tested with the smaller three-hour subset for a more compact vocabulary. Encoder outputs and codebook entries are L2-normalized before distance (Equation~\ref{eq:L2N}).

    \begin{equation}
    k^* = \arg\min_{k} \|\text{norm}(\mathbf{z}_e) - \text{norm}(\mathbf{e}_k)\|_2
    \label{eq:L2N}
    \end{equation}

    The continuous representation is replaced
    with this entry (Equation~\ref{eq:codbk}).

    \begin{equation}
    \mathbf{z}_q = \mathbf{e}_{k^*}
    \label{eq:codbk}
    \end{equation}

    Gradients bypass the non-differentiable quantization step via
    straight-through estimation, and the codebook is updated by EMA with decay $\alpha=0.99$ (Equation~\ref{eq:ema}). 

    \begin{equation}
    \mathbf{e}_k \leftarrow \alpha \mathbf{e}_k + (1-\alpha) \bar{\mathbf{z}}_e^{(k)}
    \label{eq:ema}
    \end{equation}

    where $\bar{\mathbf{z}}_e^{(k)}$ is the running mean of encoder outputs assigned to entry $k$.

    \item \textbf{Decoder:} Reconstructs the original 1295-dimensional space from $\mathbf{z}_q$ through four stages
    ($128 \rightarrow 512 \rightarrow 512 \rightarrow 1024 \rightarrow 1295$) with LayerNorm, GELU, dropout, a residual connection at the first stage, and a final linear projection without activation. The decoder output $\hat{\mathbf{x}} \in \mathbb{R}^{1295}$ approximates the original PaSST
    embedding.
\end{itemize}

\subsection{Training Objective}
The VQ-VAE is trained using a composite loss function consisting of a
reconstruction term and a combined quantization term
(Equation~\ref{eq:total_loss}):

\begin{equation}
\begin{aligned}
&\mathcal{L}_{\text{total}} = \mathcal{L}_{\text{recon}} + \lambda
\mathcal{L}_{\text{vq}} \\[6pt]
&\text{where: } \lambda = 0.1
\end{aligned}
\label{eq:total_loss}
\end{equation}

The outer weight $\lambda = 0.1$ ensures that the reconstruction term $\mathcal{L}_{\text{recon}}$ remains the dominant training signal, quantization is kept as a regulariser to prevent the codebook from collapsing in early training.
The quantization loss $\mathcal{L}_{\text{vq}}$ is a combination of three
terms (Equation~\ref{eq:vq_loss}):

\begin{equation}
\begin{aligned}
&\mathcal{L}_{\text{vq}} = \mathcal{L}_{\text{codebook}} + \beta
\mathcal{L}_{\text{commit}} + \gamma \mathcal{L}_{\text{diversity}} \\[6pt]
&\text{where: } \beta = 0.25,\quad \gamma = 0.1
\end{aligned}
\label{eq:vq_loss}
\end{equation}

Each term serves a distinct objective:

\begin{itemize}
\item \textbf{Reconstruction Loss} measures the mean squared error between
the original PaSST embedding and the decoder output
(Equation~\ref{eq:recon_loss}):

\begin{equation}
\begin{aligned}
&\mathcal{L}_{\text{recon}} = \| \mathbf{x} - \hat{\mathbf{x}} \|^2 \\[6pt]
&\text{where: } \mathbf{x} \in \mathbb{R}^{1295} \text{ is the original
input},\\
&\quad \hat{\mathbf{x}} \in \mathbb{R}^{1295} \text{ is the reconstructed
output}
\end{aligned}
\label{eq:recon_loss}
\end{equation}

\item \textbf{Codebook Loss} pulls codebook entries toward the encoder
outputs assigned to them, ensuring the vocabulary remains relevant to what
the encoder is producing (Equation~\ref{eq:codebook_loss}):

\begin{equation}
\begin{aligned}
&\mathcal{L}_{\text{codebook}} = \| \mathbf{z}_q -
\text{sg}[\mathbf{z}_e] \|^2 \\[6pt]
&\text{where: } \text{sg}[\cdot] \text{ is the stop-gradient operator, } \\
&\quad \mathbf{z}_q \text{ is the quantized representation}
\end{aligned}
\label{eq:codebook_loss}
\end{equation}

\item \textbf{Commitment Loss} encourages the encoder to commit to a
codebook entry rather than fluctuating between multiple entries
(Equation~\ref{eq:commit_loss}):

\begin{equation}
\begin{aligned}
&\mathcal{L}_{\text{commit}} = \| \text{sg}[\mathbf{z}_q] -
\mathbf{z}_e \|^2 \\[6pt]
&\text{where: } \mathbf{z}_e \in \mathbb{R}^{128} \text{ is the continuous
encoder output}
\end{aligned}
\label{eq:commit_loss}
\end{equation}

\item \textbf{Diversity Loss} encourages uniform utilization of the codebook
through entropy regularization, preventing codebook collapse
(Equation~\ref{eq:diversity_loss}):

\begin{equation}
\begin{aligned}
&\mathcal{L}_{\text{diversity}} = -\sum_{k=1}^{K} p_k \log p_k \\[6pt]
&\text{where: } p_k \text{ is the usage frequency of codebook entry } k,\\
&\quad K \in \{32, 64\}
\end{aligned}
\label{eq:diversity_loss}
\end{equation}

\end{itemize}

Training proceeds in two phases: for the first 10 epochs only
$\mathcal{L}_{\text{recon}}$ is active, allowing the encoder and decoder to
establish a meaningful representation before quantization is introduced and
preventing early codebook collapse. After epoch 10 the full
$\mathcal{L}_{\text{total}}$ is applied. Early stopping halts training when
validation loss does not improve by more than 0.0005 over 15 epochs, subject
to a minimum active token threshold of $\lfloor K/6 \rfloor$, ensuring
training is not terminated while the codebook is still forming. Following
training, a post-processing step merges tokens with cosine similarity
exceeding 0.92 and removes tokens with usage below 2\%, reassigning their
frames to the nearest active entry and renumbering sequentially.

\section{Experimental Setup}

\subsection{Model Quality Evaluation}
The following metrics assess the quality of the VQ-VAE and its learned codebook across experiments.

\textbf{Reconstruction Quality:} The primary quantitative measure of model
performance is the mean squared error between PaSST embeddings and decoder
reconstructions, assessed through Equation~\ref{eq:recon_loss}.

\textbf{Codebook Utilization:} Codebook perplexity measures how uniformly
the model uses its vocabulary, (Equation~\ref{eq:perplexity}):

    \begin{equation}
    \begin{aligned}
    &\text{Perplexity} = \exp\left(-\sum_{k=1}^{K} p_k \log p_k\right) \\[6pt]
    &\text{where: } p_k \text{ is the usage frequency of codebook entry } k
    \end{aligned}
    \label{eq:perplexity}
    \end{equation}

    Higher perplexity indicates more uniform codebook usage, meaning the model
    is using a larger portion of its vocabulary to represent the data.

\subsection{Validation Against Known Conditions}
A core objective of this work is to determine whether the learned
representations capture meaningful acoustic states without supervision. Token usage distributions and latent embeddings are examined post-hoc against known hive conditions. queen status (queenright vs.\ queenless) is used as the
evaluation label, matched to recordings via the nearest preceding inspection date from \texttt{inspections\_2021.csv}. These labels are never exposed during training.

\textbf{Token Distribution Between States:} The Jensen-Shannon Divergence
(JSD) between queenright and queenless token usage distributions quantifies
how distinctly the two conditions use the learned codebook.

\textbf{2D Latent Projection:} Latent embeddings are projected to 2D using
UMAP and t-SNE for visual inspection of cluster structure and separation
between conditions.

\textbf{Spatial Separation:} Separation in the latent space is assessed using
the silhouette score and a nearest-neighbour outlier analysis, measuring the
fraction of queenless embeddings spatially closer to the queenright cluster
than to their own condition.

\textbf{Sub-state Analysis:} Internal structure within the queenless condition
is investigated by applying $k$-means clustering ($k=3$) to queenless
embeddings in the full 128-dimensional space. Sub-states are characterised by
their dominant token and token purity, and examined through a PCA projection
computed exclusively on queenless embeddings to remove distortion from the
queenright mass.

\subsection{Temporal Analysis}
Token sequences are treated as a discrete time series to evaluate token transition, using three metrics.

\textbf{Token Transition Matrix:} A first-order transition count matrix
$C \in \mathbb{Z}^{K \times K}$ is accumulated from all consecutive token pairs $(t_i, t_{i+1})$, then row-normalised to give conditional
probabilities $P(t_{i+1} \mid t_i)$.

\textbf{Transition Entropy:} Shannon entropy of each token's outgoing distribution measures how predictable its successor is:
\begin{equation}
H_i = -\sum_{j} P(t_{i+1}{=}j \mid t_i{=}i)\,\log_2 P(t_{i+1}{=}j \mid t_i{=}i)
\label{eq:trans_entropy}
\end{equation}
The ratio $H_i / H_{\max}$, where $H_{\max} = \log_2 K_{\text{active}}$,
normalises for codebook size.

\textbf{Statistical Tests:} Two chi-squared tests assess if temporal structure is statistically significant. A goodness-of-fit test against a
uniform baseline. A second chi-squared
independence test applied to the full transition sub-matrix of active tokens observes if joint distribution $P(t_{i+1}, t_i)$ departs from the product of marginals, i.e.\ whether knowledge of the current token provides information about the next.

\subsection{Generalization to Unseen Data}
Token distribution shift between training and unseen recordings is quantified
by JSD, where values below 0.2 indicate stable generalization and values above
0.3 indicate significant distribution shift. Jaccard overlap measures the
proportion of active training tokens that remain active on the test recording.
Manifold consistency is assessed by projecting unseen embeddings through the
UMAP fitted on training data (Equations~\ref{eq:umap_high}-\ref{eq:umap_low}), placing each test point relative to its nearest neighbours in the training set so that topology can
be compared directly across seen and unseen recordings.

\begin{equation}
p_{j|i} = \exp\left(-\frac{\max(0,\, d(x_i, x_j) - \rho_i)}{\sigma_i}\right)
\label{eq:umap_high}
\end{equation}

\begin{equation}
q_{ij} = \frac{1}{1 + a \|y_i - y_j\|^{2b}}
\label{eq:umap_low}
\end{equation}

where $\rho_i$ is the distance to the nearest neighbour of $x_i$, $\sigma_i$
normalises local density, $y_i \in \mathbb{R}^3$ is the low-dimensional
position of $x_i$, and $a, b$ are fitted constants.

\section{Experimental Results}

\subsection{Experiment Configuration}
Three full-scale experiments were conducted. The baseline (E1\_baseline) uses
5 hours of training data, random seed 0, and a codebook of size 64.
E2\_small\_codebook uses 3 hours of data with a codebook of 32 to assess the
effect of vocabulary size. E1\_baseline\_seed1 repeats the baseline with a
different random seed to evaluate robustness. Prior to full-scale training,
small-scale experiments on 1 minute and 30 minutes of audio confirmed that
the VQ-VAE learns structured representations even under severely limited data.

\subsection{Model Evaluation}

Given the unsupervised nature of this work, codebook distribution and
reconstruction error are the primary quality indicators. All figures in the
following subsections refer to E1\_baseline evaluated on an unseen recording.
Table~\ref{tab:model_eval} summarises the configuration and key metrics for
each experiment.

% ── TABLE 1: Model Evaluation ──────────────────────────────────────────
\begin{table}[t]
\centering
\caption{Model Evaluation Metrics}
\label{tab:model_eval}
\resizebox{\columnwidth}{!}{%
\begin{tabular}{lcccccccc}
\hline
\textbf{Experiment} & \textbf{Training Data} & \textbf{Codebook} & \textbf{Seed} & \textbf{Epochs} & \textbf{Recon Loss} & \textbf{Perplexity} & \textbf{Active Tokens} \\
\hline
E1\_baseline        & 350k frames (5h) & 64 & 0 & 25 & 0.91 & 15.82 & 19/64 \\
E1\_baseline\_seed1 & 350k frames (5h) & 64 & 1 & 22 & 0.93 & 14.54 & 17/64 \\
E2\_small\_codebook & 210k frames (3h) & 32 & 0 & 22 & 1.30 & 16.64 & 18/32 \\
\hline
\end{tabular}%
}
\end{table}

\subsubsection{Reconstruction}
\begin{figure}[!htbp]
    \centering
    \includegraphics[width=0.8\columnwidth]{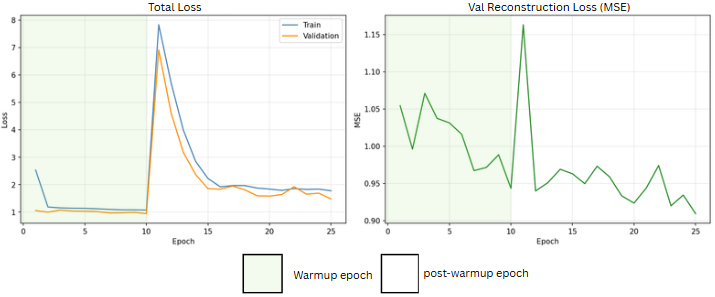}
    \caption{Loss and reconstruction curves for the baseline experiment.}
    \label{fig:recon_curv}
\end{figure}

Figure~\ref{fig:recon_curv} shows the total loss and validation reconstruction
loss for the baseline experiment. The warmup period is identifiable as the
green shaded area in the plot. During epochs 1-10, only the reconstruction
loss drives backpropagation, and the model converges to a stable
reconstruction loss. At epoch 11, the full quantization loss is introduced,
which causes a sharp spike in total loss as the vector quantizer begins
pulling outputs towards codebook entries. This spike is expected behaviour at this stage of training, arising from the discontinuity introduced by activating the quantization term. The reconstruction loss itself also spikes but recovers much more quickly. Overall loss remains at a higher plateau, but the model recovers and reconstruction continues to improve.

\begin{figure}[!htbp]
    \centering
    \includegraphics[width=\columnwidth]{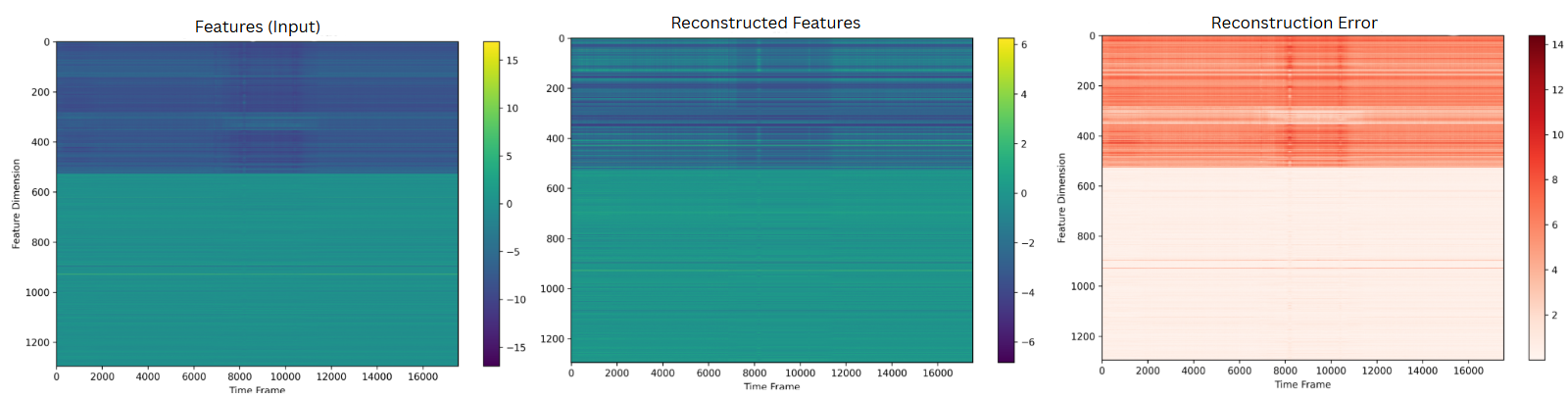}
    \caption{Reconstruction quality on the test recording.}
    \label{fig:recon_Wav}
\end{figure}

Figure~\ref{fig:recon_Wav} shows the reconstruction quality for the test
recording. The decoder output is compared against the original PaSST features.
Large error margins concentrate in the upper band of dimensions, which is the
high-activation region of the PaSST embedding space, while the lower
dimensions (approximately 700 to 1295) are reconstructed with near-zero error.
This pattern is a consequence of the discrete bottleneck and reflects correct
model behaviour rather than failure. PaSST processes audio as mel spectrogram
patches; the high-activation dimensions (0-500) encode the mid-frequency
range where bee buzzing patterns exist. The other dimensions (700-1295)
correspond to frequency content above where bees produce meaningful sound,
which is why this region shows near-zero reconstruction error. The codebook
learns one representative embedding per token, capturing the dominant pattern
across frames assigned to that token. Within-state variation is largest in the
high-activation dimensions precisely because that is where the bee signal is
richest and most variable. The concentration of error in dimensions 0-500
indicates that is where the model spends its representational capacity, which
is exactly where it should. The goal is not perfect reconstruction. A codebook
that reconstructed everything perfectly would not be performing any meaningful
compression. Fidelity loss is unavoidable when collapsing continuous embeddings
into one of a fixed set of discrete codes, particularly given the high
dimensionality of the data.

The codebook represents the colony-level acoustic state rather than preserving
individual frame variation. The remaining error reflects the deviation of each
frame from the average pattern of its assigned token, which is the expected
cost of discrete compression.

\subsubsection{Codebook}
\begin{figure}[!htbp]
    \centering
    \includegraphics[width=0.8\columnwidth]{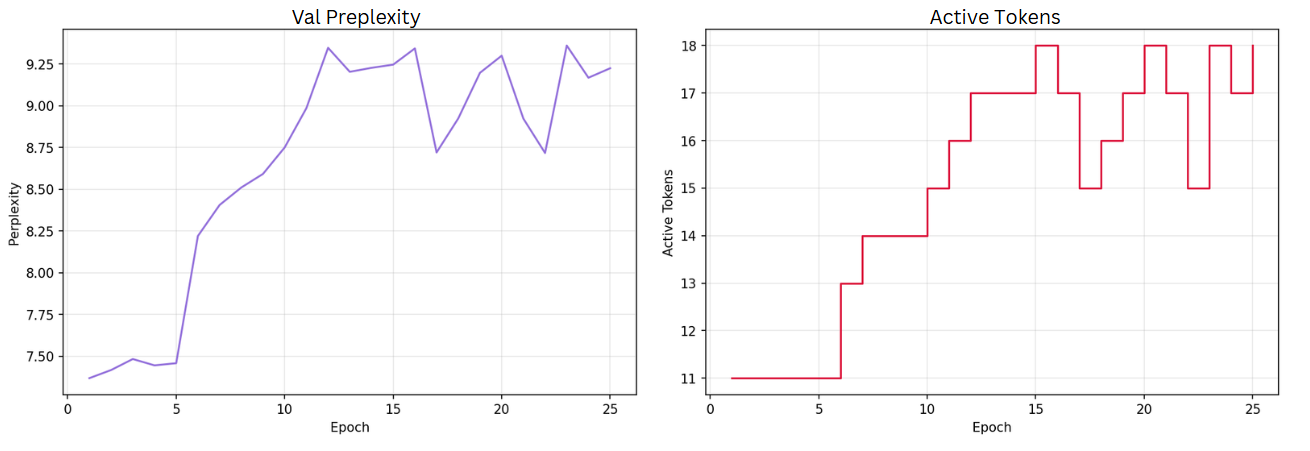}
    \caption{Codebook perplexity and active token count across training epochs.}
    \label{fig:code_wav}
\end{figure}

Figure~\ref{fig:code_wav} shows the validation perplexity and active token
count across epochs. Perplexity rises steadily from 7.5 to approximately 9.25,
reflecting that token usage grows more evenly across the codebook. Active
tokens grow from 11 at the start of training to 18 by the final epochs,
showing no codebook collapse.

\begin{figure}[!htbp]
    \centering
    \includegraphics[width=0.8\columnwidth]{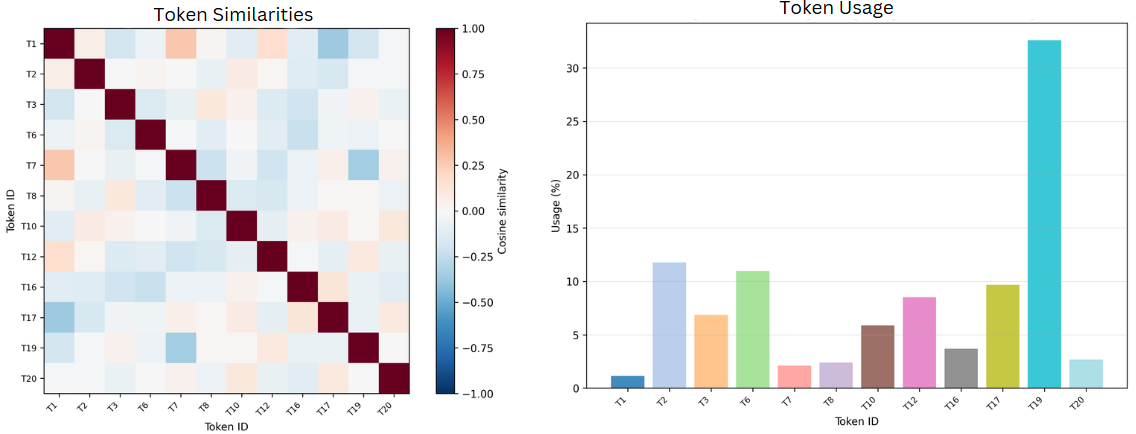}
    \caption{similarity matrix, and usage distribution.}
    \label{fig:cdbkuas}
\end{figure}

Figure~\ref{fig:cdbkuas} shows the learned codebook structure. The PCA
projection shows tokens are well separated and spread out across the space. A
similarity matrix confirms that token pairs have low cosine similarity, with
only some pairs showing moderate similarity. The usage bar chart shows that
most tokens are used at comparable frequencies, though token 19 is used
substantially more than the rest.

\subsection{State Validation}
\label{sec:state_val}
To determine whether the concentration of error in the high-activation
dimensions reflects meaningful learning, and whether the encoder captures
genuine state variation, the learned representations were evaluated against
known hive conditions. Token usage patterns were analyzed across different
states, and embeddings were projected to 2D using UMAP to observe whether
spatial separation occurred between conditions.

\subsubsection{Validation Data and Labeling}
Inference was performed on all recordings, producing token assignments for
each audio frame. queen status labels were derived from hive inspection records
using the \texttt{inspections\_2021.csv} annotation file. Each recording was
matched to the nearest preceding inspection entry for its hive, assigning
either a queenright (QR) label, denoting a colony with a functioning queen, or
a queenless (QNL) label, denoting a colony from which the queen had been
removed. Of the 326 recordings, all were successfully matched, yielding 262
queenright files (4.59M frames) and 64 queenless files (1.12M frames), a class
ratio of approximately 4:1. These labels were not used during training and are
employed solely for post-hoc validation purposes.

\subsubsection{Validation Results}
Table~\ref{tab:validation_detailed} summarises the metrics for state validation
across all experiments. Token usage separation between the two known conditions
is substantial, with Jensen-Shannon Divergence values ranging from 0.609 to
0.688. This separation was achieved without any label supervision during
training. queenright colonies consistently utilized a vocabulary of 13-16
active tokens with Shannon entropy of 2.04-2.40 bits, reflecting diverse and
distributed token usage. queenless colonies, by contrast, collapsed to only
5-6 active tokens with entropy of 1.13-1.25 bits, with a single dominant
token accounting for 56-58\% of all frames across all experiments; however,
this could be due to the limitation of queenless data compared to queenright.
Most importantly, the proportion of QNL outliers falling within the QR cluster
remains below 2\% across all experiments, indicating that the model can
generally differentiate between the QNL and QR states.

% ── TABLE 2: State Validation ───────────────────────────────────────────
\begin{table}[t]
\centering
\caption{State validation metrics across all experiments.}
\label{tab:validation_detailed}
\resizebox{\columnwidth}{!}{%
\begin{tabular}{llcccccc}
\hline
\textbf{Experiment} & \textbf{Condition} & \textbf{JSD} & \textbf{Active Tokens} & \textbf{Entropy (bits)} & \textbf{Top Token (\%)} & \textbf{Silhouette} & \textbf{QNL Outliers} \\
\hline
\multirow{2}{*}{E1\_baseline}
  & queenright & \multirow{2}{*}{0.609} & 13/64 & 2.042 & 39.04 & \multirow{2}{*}{0.046} & \multirow{2}{*}{1.57\%} \\
  & queenless  &                        &  5/64 & 1.134 & 58.00 & & \\
\hline
\multirow{2}{*}{E1\_baseline\_seed1}
  & queenright & \multirow{2}{*}{0.688} & 13/64 & 2.210 & 27.68 & \multirow{2}{*}{0.016} & \multirow{2}{*}{1.57\%} \\
  & queenless  &                        &  6/64 & 1.187 & 56.30 & & \\
\hline
\multirow{2}{*}{E2\_small\_codebook}
  & queenright & \multirow{2}{*}{0.663} & 16/32 & 2.398 & 19.94 & \multirow{2}{*}{0.188} & \multirow{2}{*}{1.70\%} \\
  & queenless  &                        &  6/32 & 1.247 & 56.45 & & \\
\hline
\end{tabular}%
}
\end{table}

Further analysis of the metrics in Table~\ref{tab:validation_detailed}
identifies how the model learned these patterns in an unsupervised manner; the
metrics each capture a distinct aspect of the learned representations. The
Jensen-Shannon Divergence (JSD) operates on the token usage distributions and
measures how different the two conditions are in terms of which codebook
entries they use and how frequently. A JSD of 0 indicates identical
distributions (no pattern differences across states) while a value of 1
indicates completely non-overlapping distributions; the 0.609-0.688 range
observed here shows that there are clearly distinct patterns between the two
states. JSD is computed over the entire condition rather than individual
frames, so it reflects a global acoustic signature rather than the behaviour of a single recording.

\begin{figure}[!htbp]
    \centering
    \includegraphics[width=1\columnwidth]{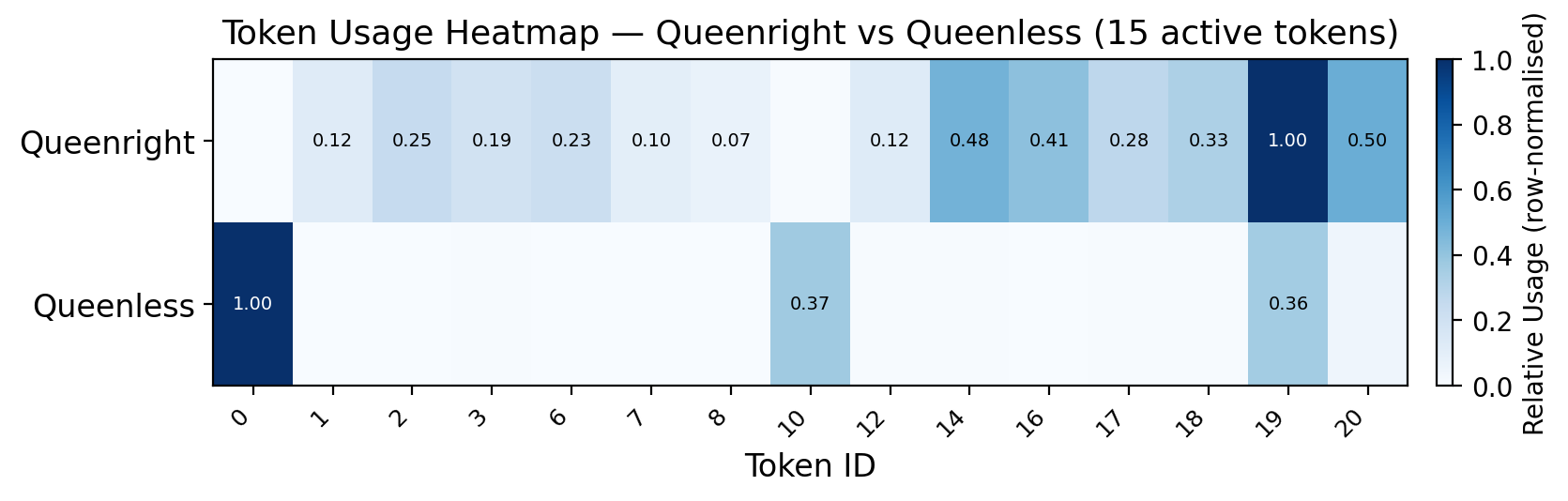}
    \caption{Baseline queen status token usage heatmap.}
    \label{fig:token_heatmap}
\end{figure}

The Shannon entropy of a condition's token distribution measures the diversity
of acoustic patterns produced: a high entropy indicates that many tokens are
used with roughly equal frequency, while a low entropy reflects concentration
on a small number of tokens. The token usage in Figure~\ref{fig:token_heatmap}
shows that queenright colonies exhibit more diverse token usage with one
dominant token while the remainder are used at a more stable frequency;
queenless colonies, in contrast, show limited token usage with only three
tokens being dominant. This further supports the lower entropy observed in
queenless colonies compared to the higher entropy of queenright colonies.

The silhouette score operates in the 128-dimensional latent space rather than
on token distributions, measuring whether individual frame embeddings cluster
more tightly with frames of their own condition than with frames of the
opposite condition. Positive values indicate that points are closer to their
own group than to the other; the moderate scores observed here (0.016-0.188)
reflect that while the two conditions are tokenwise very distinct, their latent
embeddings partially overlap in the continuous space prior to quantization.

\begin{figure}[!htbp]
    \centering
    \includegraphics[width=0.8\columnwidth]{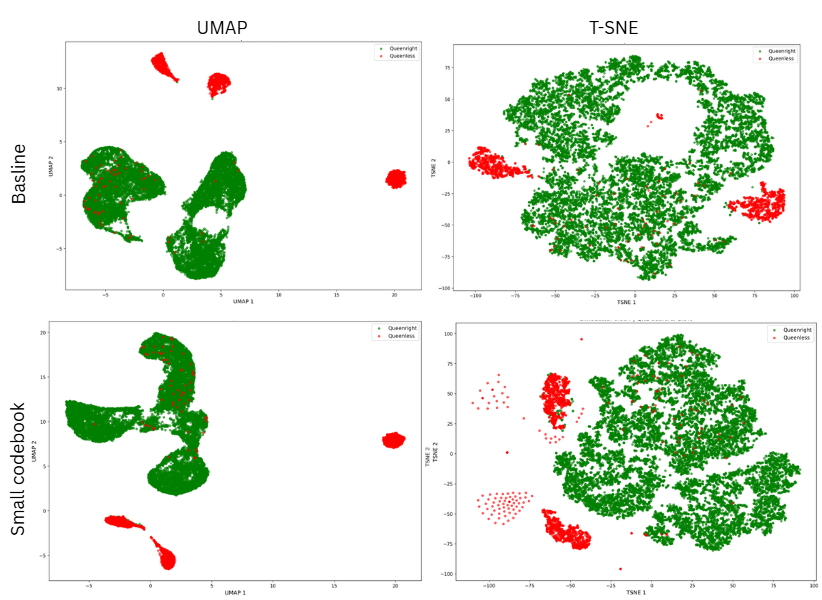}
    \caption{2D latent projections coloured by queen status.}
    \label{fig:qnstat}
\end{figure}

Finally, the QNL outlier percentage measures the fraction of queenless frame
embeddings that are spatially closer to the queenright cluster centroid than to
their own; the consistently low values across all experiments confirm that
queenless representations form a coherent and self-contained region in the
latent space, despite the class imbalance in the dataset. The embeddings are
clustered in their 128-dimensional latent space and projected into more
interpretable forms using dimensionality reduction methods. UMAP and t-SNE
projections are shown for two different experiments in
Figure~\ref{fig:qnstat}. It can be observed that in these projections,
queenright embeddings form a single connected region while queenless frames do
not form a single contiguous region but instead appear as two to three isolated
clusters separated from the main queenright mass.

\subsubsection{Sub-states}
An interesting observation from the state validation projections is that UMAP
and t-SNE form distinct clusters that are not fully connected. In t-SNE, two
clusters are clearly dominant while a third is either barely formed or sparsely
populated. To investigate this, QNL embeddings were isolated in the
128-dimensional latent space and K-means clustering ($k = 3$) was applied.
Three distinct sub-states were identified across all three experiments.

% ── TABLE 3: Sub-states ─────────────────────────────────────────────────
\begin{table}[t]
\centering
\caption{Sub-state statistics across all experiments.}
\label{tab:substates}
\resizebox{\columnwidth}{!}{%
\begin{tabular}{lcccccccccc}
\hline
 & \multicolumn{3}{c}{\textbf{Sub-state A}} & \multicolumn{3}{c}{\textbf{Sub-state B}} & \multicolumn{3}{c}{\textbf{Sub-state C}} \\
\cmidrule(lr){2-4} \cmidrule(lr){5-7} \cmidrule(lr){8-10}
\textbf{Experiment} & \textbf{Size (\%)} & \textbf{Dom. Token} & \textbf{Purity (\%)} & \textbf{Size (\%)} & \textbf{Dom. Token} & \textbf{Purity (\%)} & \textbf{Size (\%)} & \textbf{Dom. Token} & \textbf{Purity (\%)} \\
\hline
E1\_baseline        & 57.6 & T0  & 97.5 & 22.0 & T10 & 53.6 & 20.4 & T19 & 90.8 \\
E1\_baseline\_seed1 & 57.4 & T1  & 97.9 & 23.0 & T8  & 88.7 & 19.5 & T5  & 73.1 \\
E2\_small\_codebook & 57.1 & T12 & 97.7 & 21.6 & T3  & 89.3 & 21.3 & T11 & 41.9 \\
\hline
\end{tabular}%
}
\end{table}

\begin{figure}[t]
    \centering
    \includegraphics[width=0.8\columnwidth]{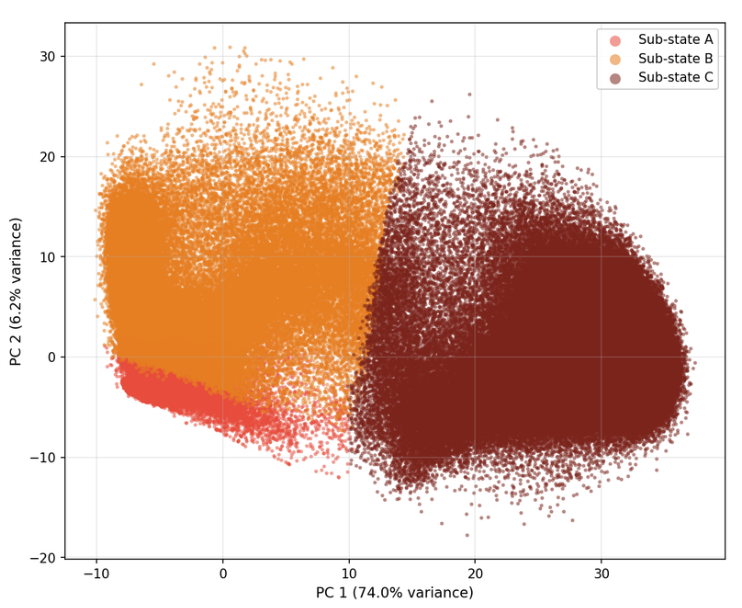}
    \caption{PCA projection of queenless embeddings coloured by sub-state.}
    \label{fig:qnl_pca}
\end{figure}

Figure~\ref{fig:qnl_pca} shows the PCA projection of the queenless embeddings
colored by sub-state for the baseline experiment. Because the projection is
solely from queenless frames, the resulting axes show the principal modes of
variation within the queenless condition. In other words, possible sub-states
within a known state can be discovered, which may reflect additional conditions
beyond the original queenless status. This setup removes the dense bias of the
queenright mass; by removing the healthy state and discovering sub-states in
the queenless condition, three spatial regions emerge, confirming that the
sub-state structure is present in the learned representations themselves.

\begin{figure}[t]
    \centering
    \includegraphics[width=0.8\columnwidth]{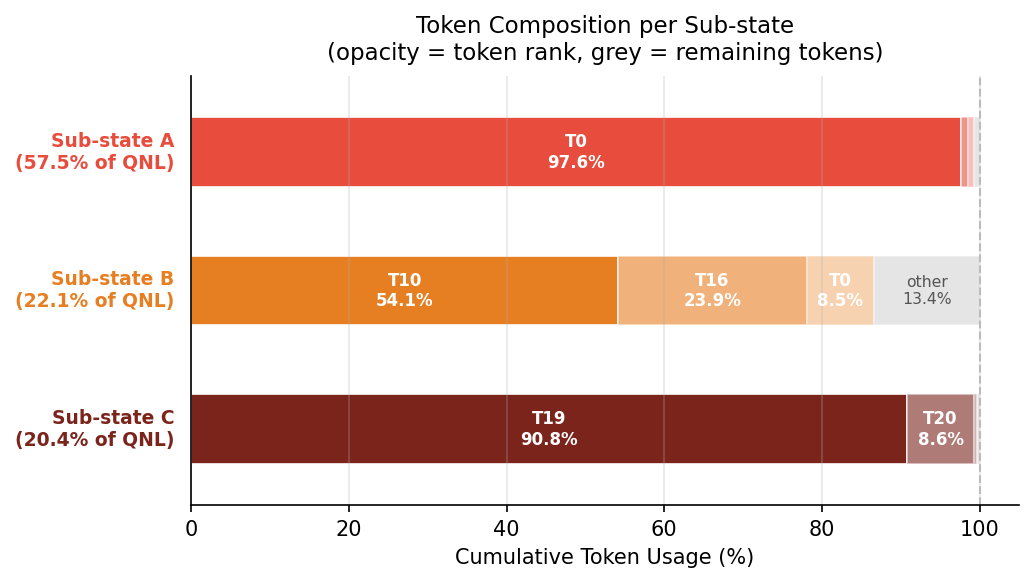}
    \caption{Token composition of queenless sub-states for the baseline experiment.}
    \label{fig:qnl_tokens}
\end{figure}

Figure~\ref{fig:qnl_tokens} shows the token composition of each sub-state.
Sub-state A accounts for 57.6\% of queenless frames and is entirely dominated
by a single token (T0 at 97.5\%), with the most acoustic concentration and
repetition in tokens of the three modes. Sub-state C comprises 20.4\% of
queenless frames and is also dominated by a single token (T19 at 90.8\%).
Sub-state B accounts for the remaining 22.0\% and exhibits a more distributed
pattern, with T10 at 53.2\% and T16 at 24.7\%, suggesting a mixture rather
than a single dominant token. The grey remainder in each bar shows the
contribution of all other tokens, which is negligible in Sub-states A and C
but reaches 13.4\% in Sub-state B.

\begin{figure}[t]
    \centering
    \includegraphics[width=0.8\columnwidth]{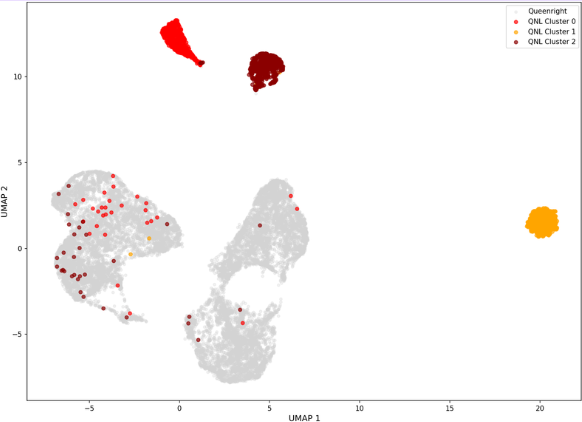}
    \caption{UMAP projection QNL sub-state.}
    \label{fig:qnl_umap}
\end{figure}

In Figure~\ref{fig:qnl_umap}, all embeddings including queenright frames are
projected through UMAP, the same baseline projection from
Figure~\ref{fig:qnstat}, with queenright frames shown in grey and queenless
frames coloured by their sub-state assignment. The UMAP projection achieves a
reasonable separation of the three sub-states, with each forming a visually
distinct region. A small number of outlier points appear in neighbouring
sub-state regions and within the queenright cluster; these are likely frames
assigned to non-dominant tokens within their sub-state, which sit closer to
the queenright manifold in the full 128-dimensional space and therefore project
into adjacent regions under UMAP's non-linear compression.

The outcome of three internally coherent sub-states within the queenless
condition suggests that queen absence does not produce a single response but a
possible set of distinct behavioural modes which the model discovered entirely
without supervision. Table~\ref{tab:substates} further supports this through
the size, dominant token, and purity of each sub-state across all three
experiments.

Based on the metrics, Sub-state A is the most consistent finding across all
experiments. Its size remains fixed at approximately 57\% of all queenless
frames, regardless of codebook size or random seed, and its purity is
consistently above 97\%, meaning that in every model trained, more than half
of all queenless audio frames collapse into a single dominant token with
near-complete uniformity. The token ID changes between experiments because IDs
are arbitrary labels assigned during training and carry no semantic meaning.

Sub-state C shows similarly high purity in E1\_baseline (90.8\%) but drops to
73.1\% in E1\_baseline\_seed1 and 41.9\% in E2\_small\_codebook. The E1 at
73\% is still acceptable; however, the drop in E2 is likely a consequence of
the smaller codebook forcing variation within fewer tokens, which dilutes the
purity of clusters. Despite this variation in purity, the size of this state
remains stable at 20-21\% across all experiments, suggesting the grouping
itself is consistent even if the token assignment is less concentrated.

Sub-state B is the least uniform of the three. Not only is there a mixture of
tokens, but the purity of the dominant token varies across models, ranging from
53.6\% in E1\_baseline to 89.3\% in E2\_small\_codebook. Its dominant token
accounts for only roughly half of its frames in two experiments, consistent
with the 13.4\% grey remainder seen in Figure~\ref{fig:qnl_tokens}, supporting
the interpretation of Sub-state B as a mixed mode rather than a single
well-defined state. However, its size also remains stable at approximately
22-23\% of queenless frames, suggesting it still consistently captures a
distinct region of the queenless space even if more heterogeneous than the
other two sub-states.

\begin{figure}[!htbp]
\centering
\includegraphics[width=0.8\columnwidth]{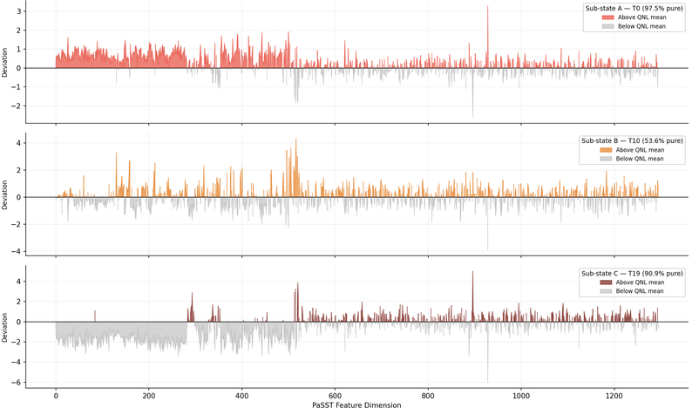}
\caption{PaSST feature deviation.}
\label{fig:passt_dev}
\end{figure}

Figure~\ref{fig:passt_dev} shows deviation of each sub-state's mean 
PaSST feature profile from the overall queenless mean. Sub-state A 
consistently activates above the queenless mean, Sub-state C is broadly suppressed below the queenless mean in the same region, suggesting lower overall energy. Sub-state B shows a mixed pattern, consistent with its interpretation as a heterogeneous behavior rather than a single well-defined state.

\subsection{Token Temporal Structure}
Table~\ref{tab:temporal_stats} summarises temporal statistics across all experiments.

% ── TABLE 4: Temporal Stats (already had resizebox, just fix table*) ────
\begin{table}[!htbp]
\centering
\caption{Temporal token sequence statistics.}
\label{tab:temporal_stats}
\resizebox{\columnwidth}{!}{%
\begin{tabular}{lccccc}
\hline
\textbf{Experiment} & \textbf{Active Tokens} & \textbf{Self-transitions} &
\textbf{Transition Entropy $H$ (bits)} & \textbf{$H/H_{\max}$} &
\textbf{Chi-squared Independence Test} \\
\hline
E1\_baseline        & 15/64 & 51\% & $2.35 \pm 0.86$ & 0.60 & $p \ll 0.001$ \\
E1\_baseline\_seed1  & 13/64 & 54\% & $2.08 \pm 0.74$ & 0.56 & $p \ll 0.001$ \\
E2\_small\_codebook  & 13/32 & 58\% & $2.42 \pm 0.58$ & 0.65 & $p \ll 0.001$ \\
\hline
\end{tabular}%
}
\end{table}

\begin{figure}[!htbp]
\centering
\includegraphics[width=0.5\columnwidth]{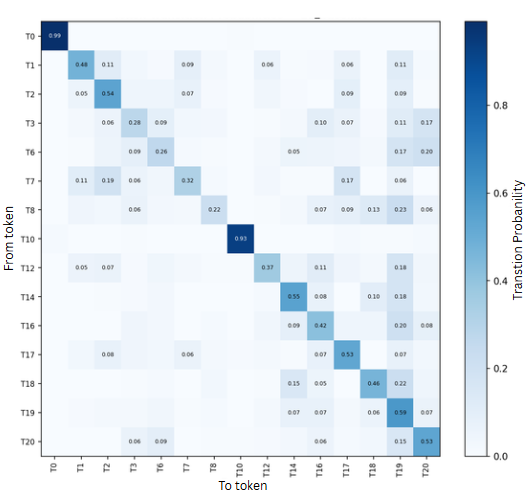}
\caption{Token transition probability matrix}
\label{fig:trans_matrix}
\end{figure}

The chi-squared independence test rejects the null hypothesis of token
independence at $p \ll 0.001$ across all experiments
($\chi^2 > 16\text{M}$, $df \leq 196$), confirming that the token
sequence carries temporal structure and is not memoryless. Every active
token exhibits non-uniform outgoing transitions ($p < 0.05$,
goodness-of-fit against uniform baseline). Outgoing transition entropy
averages $2.08$-$2.42$ bits against a maximum possible $3.70$-$3.91$
bits ($H/H_{\max} = 0.56$-$0.65$), indicating structured but
non-deterministic transitions. Two tokens exhibit markedly lower entropy:
T0 ($H = 0.10$ bits) and T10 ($H = 0.59$ bits). The transition matrix in
Figure~\ref{fig:trans_matrix} confirms this directly, T0 transitions to
itself with probability 0.99 and T10 with 0.93, while all other active
tokens distribute probability across multiple successors. These two tokens
correspond to the dominant queenless sub-states identified in
Section~\ref{sec:state_val}, suggesting their persistence is a property
of colony behavior rather than a quantization artifact.

Approximately half of all frame-to-frame transitions are self-transitions
(51-58\%), which is expected given the 23\,ms frame resolution relative
to colony-level acoustic phenomena that operate on timescales of seconds
to minutes. The remaining 42-49\% of non-self transitions are not claimed
to reflect meaningful state changes; at this resolution they are likely
dominated by quantization boundary effects and noise. The more informative
finding is that self-transition rates vary substantially across tokens
(0.28-0.99), and that when transitions do occur they follow structured
pathways rather than distributing randomly across the codebook, both
properties confirmed by the chi-squared independence test ($p \ll 0.001$)
and consistent across all three experiments.

\subsection{Unseen Data}

\textbf{Token Distribution.} Of the 19 active training tokens, 18 are 
also active in the test file, giving a Jaccard overlap of 0.947 and a JSD 
of 0.2065. The test file uses the same tokens at broadly similar relative 
frequencies, with lower absolute counts due to the smaller file size.

\begin{figure}[!htbp]
    \centering
    \includegraphics[width=\columnwidth]{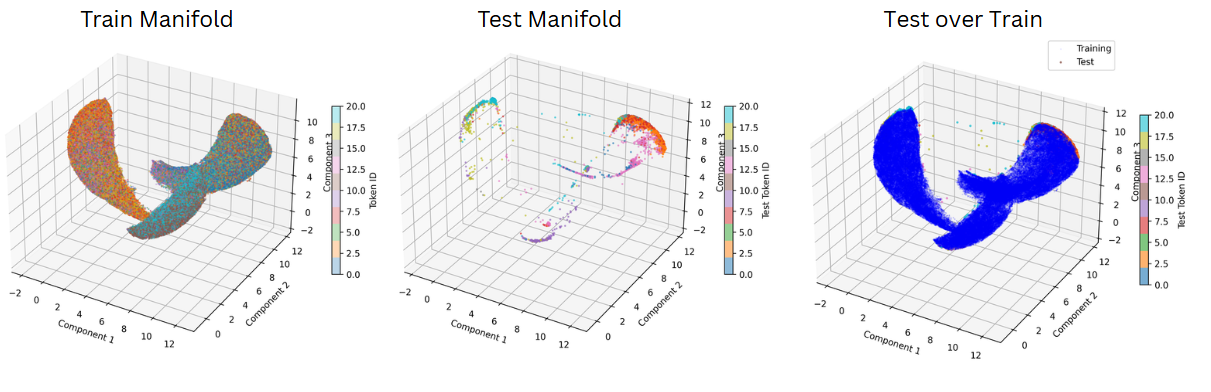}
    \caption{UMAP manifold projections for training, test, and overlay.}
    \label{fig:manifold}
\end{figure}

\textbf{Manifold Projection.} Figure~\ref{fig:manifold} shows the UMAP
manifolds for training, test, and overlay. The test manifold recovers a
similar global shape to the training manifold despite being computed from
approximately 10\% of the data (17,531 test samples vs.\ 175,336 training
samples). When overlaid, test embeddings fall primarily within one region of
the training manifold, consistent with the dominant token observed in the
token distribution, confirming that global manifold topology is preserved
under unseen input.

\section{Discussion and Limitations}

The central question of this work is whether collective honey bee buzzing
contains structured, repeatable acoustic states that can be recovered without
supervision. The results across three independent experiments consistently
support that it does, with token separation, stable sub-state structure, and
preserved manifold topology on unseen recordings all emerging without any
imposed supervision.

One question is why models such as HuBERT, wav2vec, or AVES were not used as
comparison baselines. These models carry inductive biases toward vocal
production systems: HuBERT and wav2vec target speech, AVES targets animal
vocalizations. Honey bee buzzing does not arise from such a system; it is
mechanically generated through muscle activity. Applying a vocally biased
encoder to a mechanically generated signal would be methodologically unsound
rather than informative. PaSST was selected instead as a domain-agnostic audio
encoder that carries no vocal production assumptions. Continuous embeddings
from PaSST alone capture variation but do not produce a countable, reusable
vocabulary; the vector quantization step forces the model to commit to
recurring patterns. The discretization is neither assumed to be correct nor
optimal; it is validated post-hoc by consistency, reuse, and separability. The
codebook matters here more than reconstruction; unlike work on vocal species,
generating or reconstructing similar sounds carries limited benefit when the
signal is not a direct communication means.

Whether the evaluation is sufficient is also a fair challenge. JSD, entropy,
clustering purity, and token overlap are not ground truth validation, and this
work does not claim they are. They evaluate necessary conditions for meaningful
structure rather than ground truth. Queen status is used solely as a post-hoc
analysis signal and is never exposed to the model during training, not even as
semi-supervision. It serves to validate that real state discovery has occurred.
Whether these states reflect genuine biological structure or model artefacts
cannot be fully resolved without biological annotation. However, Sub-state A
accounting for 57\% of queenless frames at over 97\% purity is robust enough
that consistent recurrence cannot plausibly be attributed to noise.

The most significant limitation is scalability. The experiments use five hours
of audio from a controlled subset of the UrBAN dataset, chosen to study
representation behaviour under interpretable conditions rather than to maximise
performance. The codebook result, while stable, may not represent the full
diversity of acoustic states across all hives, seasons, and conditions in the
full 1000 hours. The learning itself does not guarantee quality at scale; what
is demonstrated is the stability of learning under controlled conditions.

\section{Conclusion}

This work demonstrates that unsupervised discrete representation learning, using a Vector-Quantized Variational Autoencoder trained on PaSST embeddings, can recover a discrete vocabulary of acoustic states without any label supervision. 
The learned tokens are separated from queenright and queenless states, and the queenless condition decomposes into coherent sub-states, token transition analysis reveals distinct non-random sequential patterns, with statistically significant dependencies between successive tokens confirmed across all three experiments.

The broader contribution is conceptual as much as technical. This work does not claim to identify new biological behaviours; rather, it demonstrates the
capability to recover structured acoustic patterns from a non-vocal species without any prior assumptions or annotations, in a domain where most
computational approaches assume predefined labels, known behaviour categories, or vocal production structure. Honey bees are treated on their own terms,
with their signals modeled as emergent colony state rather than as 
communicative output borrowed from vocal species frameworks. Extending this approach to the full dataset and grounding the discovered sub-states through biological annotation remain the most immediate directions for future work.

Non-invasive acoustic monitoring of this kind carries practical implications 
beyond the immediate technical results. While queen loss and swarming 
detection are already established targets in hive monitoring, the 
unsupervised nature of BeeVe opens the possibility of identifying states 
beyond those currently labelled, capturing colony behaviour that inspection 
schedules and predefined classifiers may miss entirely. Early detection of 
any such state without physical inspection reduces colony disturbance and 
lowers the barrier to timely intervention, particularly for small-scale 
beekeepers. More broadly, a tool that requires no annotated data to deploy 
contributes to the movement toward species-sensitive, low-cost monitoring, 
supporting pollinator conservation at a time of accelerating colony collapse.

\bibliographystyle{amsplain}
\bibliography{ref}

\end{document}